\documentclass[aps,showpacs,amsmath,preprint,amssymb,12pt]{revtex4}
\usepackage{graphicx}

\def\al{{\alpha}}
\def\lam{{\lambda}}
\def\gam{{\gamma}}

\def\tilU{{\tilde{U}}}
\def\tU{{\tilde{U}}}

\begin{document}


\title{Rotating Black Holes on Kaluza-Klein Bubbles }

\author{Shinya Tomizawa${}^{1}$, Hideo Iguchi${}^{2}$ and Takashi Mishima${}^{2}$} 

\affiliation{${}^{1}$ Department of Mathematics and Physics, Graduate School of Science, Osaka City University, 3-3-138 Sugimoto, Sumiyoshi,
Osaka~558-8585,~Japan\\
${}^{2}$ Laboratory of Physics,~College of Science and Technology,~
Nihon University,\\ Narashinodai,~Funabashi,~Chiba 274-8501,~Japan
}

\date{\today}

\begin{abstract}
Using the solitonic solution generating techniques, 
we generate a new exact solution which describes a pair of rotating black holes on a Kaluza-Klein bubble as a vacuum solution in the five-dimensional Kaluza-Klein theory. We also investigate the properties of this solution. Two black holes with topology $S^3$ are rotating along the same direction and the bubble plays a role in holding two black holes. In static case, it coincides with the solution found by Elvang and Horowitz.
\end{abstract}

\preprint{OCU-PHYS 260}
\preprint{AP-GR 39}
\maketitle

\section{Introduction}

Solitonic solution-generating methods are powerful tools to generate exact solutions of Einstein equations. They are
mainly classified into two types.
One is called B\"acklund transformation \cite{Harrison,Neugebauer},
which is basically the technique to generate 
a new solution of the Ernst equation. The other is the inverse scattering technique,
which Belinski and Zakharov~\cite{Belinskii} developed
as an another type of solution-generating technique. Both methods have produced vacuum solutions from a certain known vacuum
solution and succeeded in generation of some
four-dimensional exact solutions. The relation between these method was discussed in Ref.~\cite{Cos} for the four dimensions.

Recently, these techniques have been used to generate five-dimensional black hole solutions.
A new stationary and axisymmetric black ring solution with rotating two sphere
was found by two of the authors~\cite{Mishima}
by applying the former solitonic solution-generating
techniques \cite{Castejon-Amenedo:1990b} to five dimensions. 
They also reproduced a black ring solution with $S^1$-rotation~\cite{Emparan:2001wn} 
 by this method \cite{MI2}
and constructed a black di-ring solution~\cite{MI3}.
As to asymptotically flat higher-dimensional black hole/ring solutions, some of solutions have been generated by using the inverse scattering method.
As an infinite number of static solutions of the five-dimensional
vacuum Einstein equations with axial symmetry,
the five-dimensional Schwarzschild solution
and the static black ring solution were reproduced \cite{Koikawa},
which gave the first example of the generation of 
a higher-dimensional asymptotically flat black hole solution 
by the inverse scattering method.
The Myers-Perry solution with single and double angular momenta
were regenerated from the Minkowski seed \cite{Tomizawa,Azuma} and
an unphysical one~\cite{Pomeransky:2005sj}, respectively.
The black ring solutions with ${ S}^2$-rotation~\cite{Tomizawa} and ${ S}^1$-rotation
~\cite{Tomizawa2} were also generated by one of the authors. Furthermore, 
Pomerasky and Sen'kov seem to succeed in generation of a new black ring solution with two angular momentum components~\cite{Pomeransky2} by the latter method. Elvang and Figueras also generated a black Saturn solution which describes a spherical black hole surrounded by a black ring~\cite{EF}.

However, from more realistic view point, we need not impose the asymptotic Minkowski spacetime toward the extra dimensions. 
In fact, higher dimensional black holes admit a variety of asymptotic structure. Kaluza-Klein black hole solutions have the spatial infinity 
with compact extra dimensions~\cite{IM,IKMT}. Black hole solutions on the Eguchi-Hanson space have the spatial infinity of topologically various lens spaces~\cite{IKMT2}. The latter black hole space-times have asymptotically and locally Minkowski structure.
In spacetimes with such asymptotic structures,  
black holes themselves have the different structures from the one 
with the asymptotically Minkowski structure. 
For instance, the Kaluza-Klein black holes~\cite{IM,IKMT} 
and the black holes on the Eguchi-Hanson space~\cite{IKMT2} admit the horizon of lens spaces in addition to ${ S}^3$. We expect that the solitonic methods also help us generate new black hole solutions which have asymptotic structures different from 
the Minkowski spacetime. Remarkably, as a vacuum solution in five-dimensional Kaluza-Klein theory, there is a static two black hole solution, which does not have even a conical singularity~\cite{EH} since a Kaluza-Klein bubble of nothing, which was first found by Witten~\cite{Witten}, plays a role in holding two black holes. In this article, we generate a new exact solution which describes a pair of rotating black holes on a Kaluza-Klein bubble by using the two different kinds of 
solution generating techniques whose relation
was discussed in \cite{Tomizawa3}. In the static case, our solution coincides with the solution found by Elvang and Horowitz~\cite{EH}. 

This article is organized as follows: In Sec.\ref{sec:solution}, we give a new solution generated by the solitonic methods. We introduce only the construction by the inverse scattering method in this section, while the other construction is briefly mentioned 
in Appendix \ref{app:Backlund}. In Sec.\ref{sec:properties}, we investigate the properties of the solution. In Sec.\ref{sec:final}, we give the summary and discussion of this article.

\section{Solutions}\label{sec:solution}
Following the techniques in the Ref~\cite{Tomizawa, Tomizawa2, Tomizawa3}, we construct a new Kaluza-Klein black hole solution. We consider the five-dimensional stationary and axisymmetric vacuum space-times which admit three commuting Killing vectors $\partial/\partial t$, $\partial/\partial \phi$ and $\partial/\partial \psi$, where $\partial/\partial t$ is a Killing vector field associated with time translation, $\partial/\partial \phi$ and $\partial/\partial \psi$ denote spacelike Killing vector fields with closed orbits. In such a space-time, the metric can be written in the canonical form as
\begin{eqnarray}
ds^2=g_{ij}dx^idx^j+f(d\rho^2+dz^2),
\end{eqnarray}
where the metric components $g_{ij}$ and the metric coefficient $f$ are functions which depend on $\rho$ and $z$ only. The metric $g_{ij}$ satisfies the supplementary condition ${\rm det}\ g_{ij}=-\rho^2$.
We begin with the following seed
\begin{eqnarray}
ds^2=-\frac{R_{\eta_2\sigma}+z-\eta_2\sigma}{R_{\eta_1\sigma}+z-\eta_1\sigma}dt^2+\frac{(R_{\eta_1\sigma}+z-\eta_1\sigma)\rho^2}{R_{\lambda\sigma}+z-\lambda\sigma}d\phi^2+\frac{R_{\lambda\sigma}+z-\lambda\sigma}{R_{\eta_2\sigma}+z-\eta_2\sigma}d\psi^2+f(d\rho^2+dz^2),\label{eq:seed1}
\end{eqnarray}
where $R_d$ is defined as $R_d:=\sqrt{\rho^2+(z-d)^2}$. The parameters $\eta_1,\eta_2$ and $\lambda$ satisfy the inequality $\eta_1<\eta_2<-1<\lambda<1$ and $\sigma>0$. 
Instead of solving the L-A pair for the seed metric (\ref{eq:seed1}), it is sufficient to consider the following metric form
\begin{eqnarray}
ds^2=-dt^2+g_2d\phi^2+g_3d\psi^2+f(d\rho^2+dz^2),\label{eq:g23}
\end{eqnarray}
where $g_2$ and $g_3$ are given by
\begin{eqnarray}
g_2=\frac{(R_{\eta_1\sigma}+z-\eta_1\sigma)^2\rho^2}{(R_{\eta_2\sigma}+z-\eta_2\sigma)(R_{\lambda\sigma}+z-\lambda\sigma)},\quad g_3=\frac{(R_{\lambda\sigma}+z-\lambda\sigma)(R_{\eta_2}+z-\eta_2\sigma)}{(R_{\eta_1\sigma}+z-\eta_1\sigma)^2}.\label{eq:g23-2}
\end{eqnarray}
Let us consider the conformal transformation of the two dimensional metric $g_{AB}\ (A,B=t,\phi)$ and the rescaling of the $\psi\psi$-component in which the determinant ${\rm det} g$ is invariant
\begin{eqnarray}
g_{0}={\rm diag} (-1,g_2,g_3)\to g'_0={\rm diag} (-\Omega,\Omega g_2,\Omega^{-2}g_3),\label{eq:conf}
\end{eqnarray} 
where $\Omega$ is the $tt$-component of the seed (\ref{eq:seed1}), i.e. 
\begin{eqnarray}
\Omega=\frac{R_{\eta_2\sigma}+z-\eta_2\sigma}{R_{\eta_1\sigma}+z-\eta_1\sigma}.
\end{eqnarray} 
Then, under this transformation, the three-dimensional metric coincides with the metric (\ref{eq:seed1}). On the other hand, as discussed in~\cite{Tomizawa3}, under this transformation the physical metric of two-solitonic solution is transformed as
\begin{eqnarray}
g=\left(
\begin{array}{@{\,}c|ccc@{\,}}
\displaystyle
g_{AB}
& 0 \\ \hline 0  &
g_3
\end{array}
\right)
\to 
g^{\prime}=\left(
\begin{array}{@{\,}c|ccc@{\,}}
\displaystyle
\Omega g_{AB}
& 0 \\ \hline 0  &
\Omega^{-2}g_3
\end{array}
\right).\label{eq:tr}
\end{eqnarray}
This is why  we may perform the transformation (\ref{eq:conf}) for the two-solitonic solution generated from the seed (\ref{eq:g23}) in order to obtain the two-solitonic solution from the seed (\ref{eq:seed1}). The generating matrix $\psi_0$ for this seed metric (\ref{eq:g23}) is computed as follows
\begin{eqnarray}
\psi_0[\lambda]={\rm diag}\left(-1,\psi_2[\lambda],\psi_3[\lambda]\right)\nonumber
\end{eqnarray}
with
\begin{eqnarray*}
& & \psi_2[\lambda]=\frac{(R_{\eta_1\sigma}+z-\eta_1\sigma+\lambda)^2(\rho^2-2z\lambda-\lambda^2)}{(R_{\eta_2\sigma}+z-\eta_2\sigma+\lambda)(R_{\lambda\sigma}+z-\lambda\sigma+\lambda)},\nonumber\\
& &\psi_3[\lambda]=\frac{(R_{\lambda\sigma}+z-\lambda\sigma+\lambda)(R_{\eta_2\sigma}+z-\eta_2\sigma+\lambda)}{(R_{\eta_1\sigma}+z-\eta_1\sigma+\lambda)^2}.
\end{eqnarray*}
Then, the two-solitonic solution is obtained as
\begin{eqnarray*}
& &g^{{\rm (phys)}}_{tt}=
-\frac{\Omega G_{tt}}{\mu_1\mu_2\Sigma},\quad
g_{t\phi}^{{\rm (phys)}}
=-g_2\frac{\Omega(\rho^2+\mu_1\mu_2)
G_{t\phi}}{\mu_1\mu_2 \Sigma},\quad
g^{{\rm (phys)}}_{\phi\phi}=
-g_2\frac{\Omega G_{\phi\phi}}{\mu_1\mu_2\Sigma},
\label{eq:gphys}
\\
& &g^{{\rm (phys)}}_{\psi\psi}=\Omega^{-2}g_3,\quad
g_{\phi\psi}^{{\rm (phys)}}=g_{t\psi}^{{\rm (phys)}}=0,
\end{eqnarray*}
where the functions $G_{tt},\ G_{t\phi},\ G_{\phi\phi}$ and $\Sigma$ are given by
\begin{eqnarray}
G_{tt}&=&-m_{01}^{(1)2}m_{01}^{(2)2}
\psi_2[\mu_1]^2\psi_2[\mu_2]^2(\mu_1-\mu_2)^2
\rho^4+m_{01}^{(1)2}m_{02}^{(2)2}g_2\mu_2^2
(\rho^2+\mu_1\mu_2)^2\psi_2[\mu_1]^2
\nonumber\\
& &+m_{01}^{(2)2}m_{02}^{(1)2}g_2 \mu_1^2
(\rho^2+\mu_1\mu_2)^2\psi_2[\mu_2]^2
-m_{02}^{(1)2}m_{02}^{(2)2}g_2^2
\mu_1^2\mu_2^2(\mu_1-\mu_2)^2\\
& &-2m_{01}^{(1)}m_{01}^{(2)}m_{02}^{(1)}m_{02}^{(2)}
g_2\psi_2[\mu_1]\psi_2[\mu_2](\rho^2+\mu_1^2)
(\rho^2+\mu_2^2)\mu_1\mu_2,
\nonumber
\end{eqnarray}
\begin{eqnarray}
G_{\phi\phi}&=&m_{01}^{(1)2}m_{01}^{(2)2}
\mu_1^2\mu_2^2(\mu_1-\mu_2)^2
\psi_2[\mu_1]^2\psi_2[\mu_2]^2
+m_{02}^{(1)2}m_{02}^{(2)2}g_2^2
(\mu_1-\mu_2)^2\rho^4\nonumber\\
& &-m_{01}^{(1)2}m_{02}^{(2)2}g_2\mu_1^2
\psi_2[\mu_1]^2(\rho^2+\mu_1\mu_2)^2
-m_{01}^{(2)2}m_{02}^{(1)2}g_2\mu_2^2
\psi_2[\mu_2]^2(\rho^2+\mu_1\mu_2)^2  \\
& &+2m_{01}^{(1)}m_{01}^{(2)}m_{02}^{(1)}m_{02}^{(2)}
g_2\mu_1\mu_2\psi_2[\mu_2]\psi_2[\mu_1]
(\rho^2+\mu_1^2)(\rho^2+\mu_2^2),
\nonumber
\end{eqnarray}
\begin{eqnarray}
G_{t\phi}&=&m_{01}^{(1)}m_{01}^{(2)2}
m_{02}^{(1)}\mu_2(\mu_1-\mu_2)
\psi_2[\mu_2]^2\psi_2[\mu_1](\rho^2+\mu_1^2)\nonumber\\
& &+m_{01}^{(1)}m_{02}^{(1)}m_{02}^{(2)2}
g_2\mu_2(\mu_2-\mu_1)
\psi_2[\mu_1](\rho^2+\mu_1^2)\nonumber\\
& &+m_{01}^{(1)2}m_{01}^{(2)}m_{02}^{(2)}
\mu_1(\mu_2-\mu_1)\psi_2[\mu_1]^2
\psi_2[\mu_2](\rho^2+\mu_2^2)\nonumber\\
&&+m_{01}^{(2)}m_{02}^{(1)2}m_{02}^{(2)}
\mu_1g_2\psi_2[\mu_2](\rho^2+\mu_2^2)(\mu_1-\mu_2),
\end{eqnarray}

\begin{eqnarray}
\Sigma&=&m_{01}^{(1)2}m_{01}^{(2)2}
\psi_2[\mu_1]^2\psi_2[\mu_2]^2(\mu_1-\mu_2)^2\rho^2
+m_{02}^{(1)2}m_{02}^{(2)2}g_2^2(\mu_1-\mu_2)^2\rho^2\nonumber\\
& &+m_{01}^{(1)2}m_{02}^{(2)2}g_2\psi_2[\mu_1]^2
(\rho^2+\mu_1\mu_2)^2
+m_{02}^{(1)2}m_{01}^{(2)2}g_2
\psi_2[\mu_2]^2(\rho^2+\mu_1\mu_2)^2\nonumber\\
& &-2m_{01}^{(1)}m_{01}^{(2)}m_{02}^{(1)}m_{02}^{(2)}
g_2\psi_2[\mu_1]\psi_2[\mu_2](\rho^2+\mu_1^2)(\rho^2+\mu_2^2).
\end{eqnarray}
Here, $\mu_1$ and $\mu_2$ are given by
\begin{eqnarray}
\mu_1(\rho,z)=\sqrt{\rho^2+(z+\sigma)^2}-(z+\sigma),\quad \mu_2(\rho,z)=\sqrt{\rho^2+(z-\sigma)^2}-(z-\sigma).
\end{eqnarray}
We should note that this three-dimensional metric $g^{{\rm (phy)}}_{ij}$ satisfies the supplementary condition ${\rm det}\ g_{ij}=-\rho^2$.  
Next, let us consider the coordinate transformation
of the physical metric such that
\begin{eqnarray}
t\rightarrow t'=t-C_1\phi, \qquad \phi
\rightarrow \phi'=\phi,
\end{eqnarray}
where $C_1$ is a constant. Under this transformation, the physical metric  becomes
\begin{eqnarray}
& &g_{tt}^{\rm (phys)}\rightarrow
g_{tt}=g_{tt}^{\rm (phys)},
\nonumber \\
& &g_{t\phi}^{\rm (phys)}\rightarrow
g_{t\phi}=g_{t\phi}^{\rm (phys)}+C_1
g_{tt}^{\rm (phys)},\label{eq:solution}\\ 
& &g_{\phi\phi}^{\rm (phys)}\rightarrow
g_{\phi\phi}=g_{\phi\phi}^{\rm (phys)}
+2C_1 g_{t\phi}^{\rm (phys)}+C_1^2g_{tt}^{\rm (phys)}.
\nonumber
\end{eqnarray}
Here, we should note that the transformed metric also satisfies the
supplementary condition ${\rm det} g=-\rho^2$. Though the metric seems to contain the four new parameters $m_{01}^{(1)},m_{01}^{(2)},m_{02}^{(1)}$ and $m_{02}^{(2)},$ it can be written only in term of the ratios
\begin{eqnarray}
\alpha:=-\frac{m_{02}^{(2)}}{2\sigma m_{01}^{(2)}},\quad \beta:=-\frac{2\sigma m_{01}^{(1)}}{m_{02}^{(1)}}. 
\end{eqnarray}
Using  the parameters $\alpha$ and $\beta$, we can write all components of the metric. The metric function  $f(\rho,z)$ takes the following form
\begin{eqnarray}
f=\frac{C_2Y_{\sigma,-\sigma}Y_{\sigma,\eta_1\sigma}}{4Y_{-\sigma,\eta_1\sigma}}\sqrt{\frac{Y_{-\sigma,\eta_2\sigma}Y_{-\sigma,\lambda\sigma}Y_{\eta_1\sigma,\eta_2\sigma}Y_{\lambda\sigma,\eta_1\sigma}Y_{\lambda\sigma,\eta_2\sigma}}{Y_{-\sigma,-\sigma}Y_{\eta_1\sigma,\eta_1\sigma}Y_{\eta_2\sigma,\eta_2\sigma}Y_{\lambda\sigma,\lambda\sigma}Y_{\sigma,\eta_2\sigma}Y_{\sigma,\lambda\sigma}Y_{\sigma,\sigma}}}\frac{\Omega Y}{(\rho^2+\mu_1\mu_2)^4\mu_1^3\mu_2\psi_2[\mu_2]^2}, \label{eq:f}
\end{eqnarray}
where $C_2$ is an arbitrary constant, $Y_{c,d}$ is defied as $Y_{c,d}:=R_cR_d+(z-c)(z-d)+\rho^2$ and the function $Y$ is given by
\begin{eqnarray}
Y&=&\rho^2[-4\beta\mu_1^2\mu_2^2\psi_2[\mu_1]\psi_2[\mu_2]+\alpha g_2(\mu_1-\mu_2)^2(\rho^2+\mu_1\mu_2)^2]^2\nonumber\\
 & &+4g_2\mu_1^2\mu_2^2(\rho^2+\mu_1\mu_2)^4(\psi_2[\mu_2]-\alpha\beta\psi_2[\mu_1])^2.\nonumber
\end{eqnarray}
We comment that the constants $\alpha$ and $\beta$ exactly coincides with the ones appeared in the B\"acklund transformation in Appendix \ref{app:Backlund}. 
To assure that the metric asymptotically approaches to ${\cal M}^{3,1}\times S^1$, where the ${\cal M}^{3,1}$ denotes the four-dimensional Minkowski spacetime and 
the $S^1$ is a Kaluza-Klein circle. The constants $C_1$ and $C_2$ are chosen as follows
\begin{eqnarray}
C_1=-\frac{2\sigma(\alpha-\beta)}{1+\alpha\beta},\quad C_2=\frac{1}{(1+\alpha\beta)^2},\quad \alpha+\beta=0,\label{eq:asymp-con}
\end{eqnarray}
to assure the regular behavior in the asymptotic region.
To avoid a singular behavior of $g_{\phi\phi}$ on the $\phi$-axis, we also need to impose the following condition on $\beta$
\begin{eqnarray}
\beta^2=-\frac{(\lambda+1)(1+\eta_2)}{(1+\eta_1)^2}.\label{eq:beta}
\end{eqnarray} 
In this article, we study the solution (\ref{eq:solution}) and (\ref{eq:f}) satisfying the conditions (\ref{eq:asymp-con}) and (\ref{eq:beta}). As mentioned later, to assure that the ADM mass is positive, we assume that the parameters $\eta_1,\eta_2$ and $\lambda$ satisfy $\beta^2<1$, i.e.
\begin{eqnarray}
(1+\eta_1)^2>-(1+\lambda)(1+\eta_2).
\end{eqnarray}


\section{Properties}\label{sec:properties}

Next, we investigate the properties of the solution satisfying the conditions (\ref{eq:asymp-con}) and (\ref{eq:beta}). In particular, we study the asymptotic structure, the geometry of two black hole horizons and a bubble and the static case.


\subsection{Asymptotic structure}
In order to investigate the asymptotic structure of the solution, let us introduce the coordinate $(r,\theta)$ defined as
\begin{eqnarray}
\rho=r\sin\theta,\quad z=r\cos \theta,
\end{eqnarray}
where $0\le\theta < 2\pi$ and $r$ is a four-dimensional radial coordinate in the neighborhood of the spatial infinity.
For the large $r\to\infty$, each component behaves as
\begin{equation}
g_{tt} \simeq -1 - \frac{\eta_1 - \eta_2 -2 - \beta^2 (\eta_1 -\eta_2 +2)}{1-\beta^2} \frac{\sigma}{r},\label{eq:a1}
\end{equation}
\begin{equation}
g_{\rho\rho}=g_{zz} \simeq 1 - \frac{\eta_1 -2 -\beta^2 (\eta_1 -\lambda +2)-\lambda}{1-\beta^2} \frac{\sigma}{r},\label{eq:a2}
\end{equation}
\begin{equation}
g_{t\phi} \simeq -\frac{2\sigma^2\beta(2\eta_1-\eta_2-\lambda-2-\beta^2(2\eta_1-\eta_2-\lambda+2))\sin^2\theta}{(1-\beta^2)^2 r},   \label{eq:a3}
\end{equation}
\begin{equation}
g_{\phi\phi} \simeq r^2 \sin^2 \theta \left(1-\frac{\eta_1-\lambda-2-\beta^2 (\eta_1-\lambda +2)}{1-\beta^2} \frac{\sigma}{r}\right),\label{eq:a4}
\end{equation}
\begin{equation}
g_{\psi\psi} \simeq 1 + \frac{\sigma(\eta_2 -\lambda)}{r}.\label{eq:a5}
\end{equation}
Hence, the leading order of the metric takes the form
\begin{eqnarray}
ds^2\simeq -dt^2+dr^2+r^2(d\theta^2+\sin^2\theta d\phi^2)+d\psi^2.
\end{eqnarray}
Therefore, the space-time asymptotically has the structure of the direct product of the four-dimensional Minkowski space-time and $S^1$. The $S^1$ at infinity is parameterized by $\psi$ and the size $\Delta\psi$ is given in \ref{sec:rod}.

\subsection{Mass and angular momentums}
Next, we compute the total mass and the total angular momentum of the space-time. It should be noted that since the asymptotic structure is ${\cal M}^{3,1}\times S^1$, the ADM mass and angular momentum are given by the surface integral over the spatial infinity with the topology of $S^2\times S^1$. In order to compute these quantities, we introduce asymptotic Cartesian coordinates $(x,y,z,\psi)$, where $x=\rho\cos\phi$ and $y=\rho\sin\phi$. Then, the ADM mass and angular momentums are given by
\begin{eqnarray}
M_{\rm ADM}=\frac{1}{16\pi}\int_{S^2\times S^1}H^{0\alpha0j}{}_{,\alpha}dS_j,
\end{eqnarray}
\begin{eqnarray}
J^{\mu\nu}=\frac{1}{16\pi}\int_{S^2\times S^1}\left(x^\mu H^{\nu\alpha 0 j}{}_{,\alpha}-x^\nu H^{\mu\alpha 0 j}{}_{,\alpha}+H^{\mu j 0 \nu}-H^{\nu j 0 \mu}\right)dS_j,
\end{eqnarray}
respectively. Here $H^{\mu\alpha\nu\beta}$ is defined by
\begin{eqnarray}
H^{\mu\alpha\nu\beta}:=-(\bar h^{\mu\nu}\eta^{\alpha\beta}+\bar h^{\alpha\beta}\eta^{\mu\nu}-\bar h^{\alpha\nu}\eta^{\beta\mu}-\bar h^{\beta\mu}\eta^{\alpha\nu}),
\end{eqnarray}
where
\begin{eqnarray}
\bar h_{\mu\nu}:=h_{\mu\nu}-\frac{1}{2}h^{\alpha}{}_{\alpha}\eta_{\mu\nu},
\end{eqnarray}
and $\eta_{\mu\nu}$ is the five-dimensional flat metric.
The Latin index $j$ runs $x,y,z$ and $\psi$ and the Greek indeces $\mu,\nu,\alpha$ and $\beta$ label $t,x,y,z$ and $\psi$.
Then, the ADM mass of the solution is computed as\begin{eqnarray}
{\cal M}_{ADM}=\frac{\sigma(4-2\eta_1+\eta_2+\lambda+\beta^2(4+2\eta_1-\eta_2-\lambda))}{4(1-\beta^2)}\Delta\psi.
\end{eqnarray}
The nonzero component of the angular momentum becomes
\begin{eqnarray}
J&=&J^{xy}=-\frac{2\beta\sigma^2(2-2\eta_1+\eta_2+\lambda+\beta^2(2+2\eta_1-\eta_2-\lambda))}{(1-\beta^2)^2}\Delta\psi.
\end{eqnarray}
It is should be noted that the ADM mass is non-negative when $\beta^2<1$.



\subsection{Black holes and bubble}\label{sec:rod}
\begin{figure}[htbp]
\begin{center}
\includegraphics[width=0.6\linewidth]{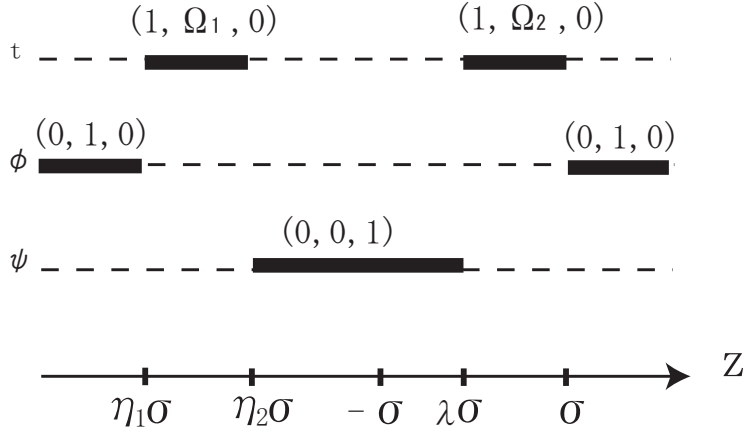}
\end{center}
\caption{Rod structure of rotating black holes on a Kaluza-Klein bubble. The finite timelike rods $[\eta_1\sigma,\eta_2\sigma]$ and $[\lambda\sigma,\sigma]$ correspond to rotating black holes with angular velocities $\Omega_1$ and $\Omega_2$, respectively. The finite spacelike rod $[\eta_2\sigma,\lambda\sigma]$ denotes a Kaluza-Klein bubble, where Kaluza-Klein circles shrink to zero.}\label{fig:rod}
\end{figure}

Here, for the solution, we consider the rod structure developed by Harmark~\cite{Harmark} and Emparan and Reall~\cite{weyl}. 
The rod structure at $\rho=0$ is illustrated in FIG.\ref{fig:rod}.
(i) The finite timelike rod $[\eta_1\sigma,\eta_2\sigma]$ and $[\lambda\sigma,\sigma]$ denote the locations of black hole horizons. These timelike rods have directions $v_1=(1,\Omega_1,0)$ and $v_2=(1,\Omega_2,0)$, where $\Omega_1$ and $\Omega_2$ mean angular velocities of the horizons. These are given by
\begin{equation}
\Omega_1 = \frac{-\beta(1-\beta^2)}
             {[1-\eta_1+\beta^2(1+\eta_1)]^2\sigma},
\end{equation} 
for $\eta_1\sigma < z < \eta_2 \sigma$ and
\begin{equation}
 \Omega_2 = \frac{-\beta(1-\beta^2)((1-\eta_2)(1-\lambda)+(1-\eta_1)^2)}
                 {4[(1-\eta_1)^2+\beta^2(1-\eta_2)(1-\lambda)]\sigma},
\end{equation} 
for $\lambda\sigma < z < \sigma.$ Here, it should be noted that $\Omega_1$ and $\Omega_2$ have the same signature. Therefore, two black holes are rotating along the same direction. (ii) The finite spacelike rod $[\eta_2\sigma,\lambda\sigma]$ 
which corresponds to a Kaluza-Klein bubble has the direction $v=(0,0,1)$. In order to avoid conical singularity for $z\in[\eta_2\sigma,\lambda\sigma]$ and $\rho=0$, $\psi$ has the periodicity of

\begin{eqnarray}
\frac{\Delta\psi}{2\pi} &=& \lim_{\rho\rightarrow \infty} \sqrt{\frac{\rho^2g_{\rho\rho}}{g_{\psi\psi}}} \nonumber\\
            &=& \frac{2\sigma}{1-\beta^2}
                            \left(\frac{\eta_1-1}{\eta_1+1}\right)
             \sqrt{\frac{(\lambda-\eta_1)(\lambda-\eta_2)(\eta_2+1)}{\eta_2-1}}
             \left(1-\frac{(\eta_1+1)^2(\eta_2-1)}{(\eta_1-1)^2(\eta_2+1)}
                                               \beta^2\right) \nonumber \\
            &=& 2\sigma\frac{(\eta_1+1)((\eta_1-1)^2+(\lambda+1)(\eta_2-1))}
                           {(\eta_1-1)((\eta_1+1)^2+(\lambda+1)(\eta_2+1))}
             \sqrt{\frac{(\lambda-\eta_1)(\lambda-\eta_2)(\eta_2+1)}{\eta_2-1}}.
\end{eqnarray}
(iii) The semi-infinite space-like rods $[-\infty,\eta_1\sigma]$ and $[\sigma,\infty]$ have the direction $v=(0,1,0)$. In order to avoid conical singularity,
$\phi$ has the periodicity of
\begin{equation}
\Delta\phi= 2\pi.
\end{equation}

Here, we write the induced metrics of the event horizons and the bubble. 
For $\eta_1\sigma<z<\eta_2\sigma$, the induced metric 
on the surface with constant $t$
becomes
\begin{eqnarray}
& &g_{\phi\phi}=\frac{4\sigma^2(z^2-\sigma^2)(z-\eta_1\sigma)(\lambda\sigma-z)(1-\eta_1+\beta^2(1+\eta_1))^4}{(1-\beta^2)^2(\sigma^2[(\eta_1-1)^2(z+\sigma)+\beta^2(1+\eta_1)^2(\sigma-z)]^2+4\beta^2k)},
\end{eqnarray}
\begin{eqnarray}
g_{\psi\psi}=\frac{z-\eta_2\sigma}{z-\lambda\sigma},
\end{eqnarray}
\begin{eqnarray}
 & &g_{zz}=\frac{(\eta_2-\eta_1)(\lambda-\eta_1)(\sigma^2[(\eta_1-1)^2(z+\sigma)+\beta^2(1+\eta_1)^2(\sigma-z)]^2+4\beta^2k)}{(1-\beta^2)^2(\eta_1^2-1)^2(z^2-\sigma^2)(z-\eta_1\sigma)(\eta_2\sigma-z)},
\end{eqnarray}
where the function $k(\rho,z)$ is defined as
\begin{eqnarray}
k(\rho,z):=(z-\eta_1\sigma)^2(\eta_2\sigma-z)(\lambda\sigma-z).
\end{eqnarray}
Since the $\psi$ circles shrink to zero at $z=\eta_2\sigma$ and $\phi$ circles shrink to zero at $z=\eta_1\sigma$, the spatial cross section of this black hole horizon is topologically $S^3$.
The area of the event horizon is 
\begin{eqnarray}
 A_1 &=& 16 \pi^2 \sigma^3\frac{(1+\eta_1)^2(\eta_2-\eta_1)^{3/2}(\lambda-\eta_1)
 }{(1-\eta_1)^2} \sqrt{\frac{(\lambda-\eta_2)(\eta_2+1)}{(\eta_2-1)}}
 \nonumber \\ &&\times
\frac{((1-\eta_1)^2-(1+\lambda)(1-\eta_2))(1-\eta_1^2-(1+\lambda)(1+\eta_2))^2}{((1+\eta_1)^2+(1+\lambda)(1+\eta_2))^3}.
\end{eqnarray}

For $\lambda\sigma<z<\sigma$, the induced metric takes the following form
\begin{eqnarray}
g_{\phi\phi}=\frac{16\sigma^2(\sigma-z)(z+\sigma)(z-\eta_1\sigma)(z-\eta_2\sigma)[(-1+\eta_1)^2+\beta^2(-1+\eta_2)(-1+\lambda)]^2}{(1-\beta^2)^2[4\sigma^2(-1+\eta_1)^4(z-\eta_2\sigma)(z-\lambda\sigma)+\beta^2h]},
\end{eqnarray}
\begin{eqnarray}
g_{\psi\psi}=\frac{z-\lambda\sigma}{z-\eta_2\sigma},
\end{eqnarray}
\begin{eqnarray}
g_{zz}&=&\frac{4\sigma^2(\eta_1-1)^4(z-\eta_2\sigma)(z-\lambda\sigma)+\beta^2h}{(1-\beta^2)^2(\eta_1-1)^2(\eta_2-1)(\lambda-1)(z-\eta_1\sigma)(z-\lambda\sigma)(\sigma^2-z^2)},
\end{eqnarray}
where the function $h$ is given by
\begin{eqnarray}
h(\rho,z)&:=&(z-\eta_1\sigma)^2[(\sigma-z)(-1+\eta_1)^2-(\eta_2-1)(\lambda-1)(z+\sigma)]^2.
\end{eqnarray}
Since the $\psi$ circles shrink to zero at $z=\lambda\sigma$ and $\phi$ circles shrink to zero at $z=\sigma$, the spatial cross section of this black hole horizon is also topologically $S^3$.
The area of this event horizon is 
\begin{eqnarray}
 A_2 &=& 32 \pi^2 \sigma^3\frac{(1+\eta_1)^3
 \sqrt{(\lambda-1)(\lambda-\eta_1)(\lambda-\eta_2)(1+\eta_2)}}
      {(\eta_1-1)^2(1-\eta_2)} \nonumber \\ &&\times
\frac{((1-\eta_1)^2-(1+\lambda)(1-\eta_2))((1-\lambda^2)(1-\eta_2^2)-(1-\eta_1^2)^2)}{((1+\eta_1)^2+(1+\lambda)(1+\eta_2))^3}.
\end{eqnarray}

For $\eta_2\sigma<z<\lambda\sigma$, the induced metric on the bubble can be written in the form 
\begin{eqnarray}
g_{\phi\phi}&=&\frac{4(1-c\beta^2)^2(z-\eta_1\sigma)(z-\eta_2\sigma)(\sigma^2-z^2)(z-\lambda\sigma)}{4\beta^2 d^2(z-\lambda\sigma)(z-\eta_1\sigma)^2+(z-\eta_2\sigma)(z+\sigma+\beta^2c(\sigma-z))^2}\nonumber\\
            & &+\frac{16\beta^2 p}{(1-\beta^2c)^2(\sigma^2-z^2)(z-\eta_1\sigma)}
\end{eqnarray}
\begin{eqnarray}
g_{zz}=-\frac{\sigma^2(\lambda-\eta_1)(\lambda-\eta_2)[(-1+\eta_1)^2(1+\eta_2)-\beta^2(1+\eta_1)^2(-1+\eta_2)]^2}{(1-\beta^2)^2(-1+\eta_1^2)^2(-1+\eta_2^2)(z-\eta_2\sigma)(z-\lambda\sigma)},
\end{eqnarray}
where the function $p$ is given by
\begin{eqnarray}
p(\rho,z):&=&(z-\eta_2\sigma)\left[(z+\sigma+\beta^2c(\sigma-z))^2
+\frac{4\beta^2d^2(z-\lambda\sigma)(z-\eta_1\sigma)^2}{(z-\eta_2\sigma)}\right]\nonumber\\
&\times&\Biggl[\frac{\sigma}{1-\beta^2}-\frac{(z+\sigma)}{(z-\eta_2\sigma)(z+\sigma+\beta^2
c(\sigma-z))^2+4\beta^2d^2(z-\lambda\sigma)(z-\eta_1\sigma)^2}q\Biggr]^2,
\end{eqnarray}
with
\begin{eqnarray}
q(\rho,z):&=&\frac{2\sigma
d(z-\lambda\sigma)(z-\eta_1\sigma)^2(1+\beta^2c)}{\sigma+z}\nonumber\\
&&-(z-\eta_2\sigma)\left(1+\beta^2c\frac{\sigma-z}{z+\sigma}\right)
\left(-\frac{\sigma^2(1+\eta_1)^2}{1+\eta_2}+d\frac{(z-\lambda\sigma)(z-\eta_1\sigma)^2}
{(z-\eta_2\sigma)}\right),
\end{eqnarray}
\begin{eqnarray}
c=\frac{(1+\eta_1)^2(-1+\eta_2)}{(-1+\eta_1)^2(1+\eta_2)},\quad d=\frac{-1+\eta_2}{(-1+\eta_1)^2}.
\end{eqnarray}
The $\psi$ circle vanishes for $z\in [\eta_2\sigma,\lambda\sigma],\rho=0$, which means that there exists a Kaluza-Klein bubble in this region. Since the $\phi$ circle does not vanish at $z=\eta_2\sigma$ and $z=\lambda\sigma$, this bubble on the time slice is topologically a cylinder $S^1\times R$. Therefore, there exist a Kaluza-Klein bubble between two rotating black holes with topology of $S^3$. 
The proper distance between the two black holes is
\begin{equation}
 s=\pi \sigma \frac{(\eta_1+1)}{(\eta_1-1)}
    \sqrt{\frac{(\eta_2+1)(\lambda-\eta_1)(\lambda-\eta_2)}{(\eta_2-1)}}
   \frac{((1-\eta_1)^2-(1+\lambda)(1-\eta_2))}{((1+\eta_1)^2+(1+\lambda)(1+\eta_2))}.
\end{equation}
The Kaluza-Klein bubble is significant to keep the balance of two black holes and
achieve the solution without any strut structures and singularities.
This property resembles that of the solution given by Elvang and Horowitz~\cite{EH}. In next  subsection, we will show that the solution coincides with it in static case.

\subsection{Static case}
Finally, let us consider the static case, which can be obtained by the choice of the parameter $\lambda= -1$. Then, from Eq. (\ref{eq:beta}) we see that $\beta$ vanishes. Let us define the parameters $a,b$ and $c$ as
\begin{eqnarray}
a=\frac{2-\lambda-\eta_2}{2}\sigma,\quad b=\frac{\lambda-\eta_2}{2}\sigma,\quad c=\frac{\lambda+\eta_2-2\eta_1}{2}\sigma.
\end{eqnarray}
It should be noted that $\lambda=-1$ is equal to the condition $\sigma=(a-b)/2$.
Furthermore, let us shift an origin of the $z$-coordinate such that $z\to\tilde z:=z-(\eta_2+\lambda)/2$.
Then, we obtain the metric
\begin{eqnarray}
ds^2&=&-\frac{(R_b-(\tilde z-b))(R_{-c}-(\tilde z+c))}{(R_a-(\tilde z-a))(R_{-b}-(\tilde z+b))}dt^2+(R_a-(\tilde z-a))(R_{-c}-(\tilde z+c))d\phi^2\nonumber\\
   & &+\frac{R_{-b}-(\tilde z+b)}{R_b-(\tilde z-b)}d\psi^2+\frac{Y_{a,-c}Y_{b,-b}}{4R_{a}R_{b}R_{-b}R_{-c}}\sqrt{\frac{Y_{a,b}Y_{-b,-c}}{Y_{a,-b}Y_{b,-c}}}\frac{R_a-({\tilde z}-a)}{R_{-c}-({\tilde z}+c)}(d\rho^2+d\tilde z^2),
\end{eqnarray}
where the coordinate $z$ in the definition of $R_d$ is replaced with $\tilde z$. This coincides with the solution obtained by Elvang and Horowitz~\cite{EH}, which describes non-rotating black holes on the Kaluza-Klein bubble.

\section{Summary and Discussion}\label{sec:final}

Using the solitonic solution generating methods, we generated a new exact solution which describes a pair of rotating black holes on a Kaluza-Klein bubble as a vacuum solution in the five-dimensional Kaluza-Klein theory. We also investigated the properties of this solution, particularly, its asymptotic structure, the geometry of the black hole horizons and the Kaluza-Klein bubble and the limit of static case.
The asymptotic structure is the $S^1$ bundle over the four-dimensional Minkowski space-time. Two black holes have the topological structure of $S^3$ and the bubble is topologically $S^1\times R$. The solution describes the physical situation such that two black holes have the angular velocity of the same direction and the bubble plays a role in holding two black holes. In the static case, it coincides with the solution found by Elvang and Horowitz.

We comment on the impossibility of the arbitrarily close spinning black holes of this solution. The reason is the inevitability of the closed timelike curves around the $z$-axis for any $\beta$ when we chose $\eta_2>-1$ to realize $\eta_2=\lambda$.

In this article, we concentrated on the black hole solution with a single angular momentum component. 
The investigation on the solution with two angular momentum components is enormously challenging.
In general, the inverse scattering method can generate a solution with two angular momentum components. However, as discussed in Ref. \cite{Tomizawa,Tomizawa2}, such a solution generated from our seed would have singular behavior on an axis due to the issues on the normalization. In order to obtain a solution with two angular momentum components, we need change our seed into another seed which does not satisfy the condition ${\rm det}g_{ij}=-\rho^2$. We will give such a solution in our future article.

\section*{Acknowledgements}
We thank Ken-ichi~Nakao for continuous encouragement.
This work is partially supported by Grant-in-Aid for Young Scientists (B)
(No. 17740152) from Japanese Ministry of Education, Science,
Sports, and Culture.

\appendix
\section{Solution by B\"acklund transformation}
\label{app:Backlund}
In this appendix we briefly 
present the solution obtained by the B\"acklund transformation
which was developed to apply the five dimensional case \cite{Mishima}.

The metric of the solitonic solution can be written in the 
following form
\begin{eqnarray}
ds^2 &=&e^{-T}\left[
       -e^{S}(dt-\omega d\phi)^2
       +e^{-S}\rho^2(d\phi)^2 
+e^{2\gamma-S}\left(d\rho^2+dz^2\right) \right]
  +e^{2T}(d\psi)^2.
  \label{MBmetric}
\end{eqnarray}
The function $T$ is derived from
the seed metric (\ref{eq:seed1}) as
\begin{equation}
T=\tilde{U}_{\lambda\sigma} -\tilde{U}_{\eta_2 \sigma},
\end{equation}
where the function $\tilU_{d}$ is defined as $\tilU_{d}:=\frac{1}{2}\ln\left[\,R_{d}+(z-d)\,\right]$.
The other metric functions for the five-dimensional metric 
 (\ref{MBmetric}) are obtained
by using the formulas shown by \cite{Castejon-Amenedo:1990b}, 
\begin{eqnarray}
e^{S}&=&e^{S^{(0)}}\frac{A}{B} ,  \label{e^S} \\
\omega&=&2\sigma e^{-S^{(0)}}\frac{C}{A}-C_1 \label{omega} ,    \\
e^{2\gamma}&=&C_2(x^2-1)^{-1}A
                e^{2\gamma'}, \label{e_gamma}
\end{eqnarray}
where $C_1$ and $C_2$ are constants and
$A$, $B$ and $C$ are given by
\begin{eqnarray}
&&A:=(x^2-1)(1+ab)^2-(1-y^2)(b-a)^2 \,, \\
&&B:=[(x+1)+(x-1)ab]^2+[(1+y)a+(1-y)b]^2 \,, \\
&&C:=(x^2-1)(1+ab)[(1-y)b-(1+y)a]  
+(1-y^2)(b-a)[x+1-(x-1)ab]\,,
\end{eqnarray}
and $x$ and $y$ are the prolate-spheroidal coordinates:
$\rho= \sigma\sqrt{(x^2-1)(1-y^2)}, z=\sigma x y$.
Here the function $S^{(0)}$ is a seed function which can be derived from
the seed metric (\ref{eq:seed1}) as
\begin{eqnarray}
S^{(0)} &=& \tilde{U}_{\lambda\sigma} -2 \tilde{U}_{\eta_1 \sigma}
              +\tilde{U}_{\eta_2 \sigma}. 
\end{eqnarray}
The functions $a$ and $b$, which are auxiliary potential to obtain 
the new Ernst potential for the seed by the transformation,
are given by
\begin{eqnarray}
a
&=& {\al}\cdot
   \frac{e^{2U_{\sigma}}+e^{2\tilU_{\lam\sigma}}}
        {e^{\tilU_{\lam\sigma}}}\cdot
   \frac{e^{2U_{\sigma}}+e^{2\tilU_{\eta_2\sigma}}}
        {e^{\tilU_{\eta_2\sigma}}}\cdot
   \left(\frac{e^{\tilU_{\eta_1\sigma}}}
        {e^{2U_{\sigma}}+e^{2\tilU_{\eta_1\sigma}}}\right)^2,
          \\
b
&=& \beta\cdot
   \frac{e^{\tilU_{\lam\sigma}}}
        {e^{2U_{-\sigma}}+e^{2\tilU_{\lam\sigma}}}\cdot
   \frac{e^{\tilU_{\eta_2\sigma}}}
        {e^{2U_{-\sigma}}+e^{2\tilU_{\eta_2\sigma}}}\cdot
   \left(\frac{e^{2U_{-\sigma}}+e^{2\tilU_{\eta_1\sigma}}}
        {e^{\tilU_{\eta_1\sigma}}}\right)^2.
\end{eqnarray}
where the function $U_{d}$ is defined as $U_{d}:=\frac{1}{2}\ln\left[\,R_{d}-(z-d)\,\right]$.
In addition the function $\gamma'$ is obtained as
\begin{eqnarray}
\gam' &=& \gam'_{\sigma,\sigma}+\gam'_{-\sigma,-\sigma}
+\gam'_{\lambda\sigma,\lambda\sigma}+\gam'_{\eta_1\sigma,\eta_1\sigma}
+\gam'_{\eta_2\sigma,\eta_2\sigma}
 \nonumber \\
&&-2\gam'_{\sigma,-\sigma}
+\gam'_{\sigma,\lambda\sigma}
-2\gam'_{\sigma,\eta_1\sigma}+\gam'_{\sigma,\eta_2\sigma}
-\gam'_{-\sigma,\lambda\sigma}
+2\gam'_{-\sigma,\eta_1\sigma}-\gam'_{-\sigma,\eta_2\sigma} \nonumber \\
&&-\gam'_{\lambda\sigma,\eta_1\sigma} -\gam'_{\lambda\sigma,\eta_2\sigma} 
-\gam'_{\eta_1\sigma,\eta_2\sigma} ,
\end{eqnarray}
where
\begin{equation}
\gam'_{cd}=\frac{1}{2}\tU_{c}+\frac{1}{2}\tU_{d}
           -\frac{1}{4}\ln [R_cR_d+(z-c)(z-d)+\rho^2]. \label{gam'}
\end{equation}


\end{document}